# Ridge Regression Estimated Linear Probability Model Predictions of N-glycosylation in Proteins with Structural and Sequence Data

Rajaram Gana[†], Swagata Naha[†], Raja Mazumder[♀], Radoslav Goldman[‡,†], and Sona Vasudevan[†,✉]

[†] Department of Biochemistry and Molecular & Cellular Biology, Georgetown University, Washington D.C., USA

[‡] Department of Oncology, Lombardi Comprehensive Cancer Center, Georgetown University, Washington D.C., USA

[♀] Department of Biochemistry, George Washington University, Washington, D.C., USA

[✉] Corresponding author[1] (contact email: sv67@georgetown.edu)

## Abstract

Absent experimental evidence, a robust methodology to predict the likelihood of N-glycosylation in human proteins is essential for guiding experimental work. Based on the distribution of amino acids in the neighborhood of the NxS/T sequon (N-site); the structural attributes of the N-site that include Accessible Surface Area, secondary structural elements, main-chain phi-psi, turn types; the relative location of the N-site in the primary sequence; and the nature of the glycan bound, the ridge regression estimated linear probability model is used to predict this likelihood. This model yields a Kolmogorov-Smirnov (Gini coefficient) statistic value of about 74% (89%), which is reasonable.

[1]Acknowledgments: We acknowledge help from various resources during the data collection phase of this work. We thank Marina Zhuravleva of PDB, Roman Laskowski of PDBSUM, and Ahmad Shander of ASAVIEW for their help. We thank Elliott Crooke, Professor and Chair, Department of Biochemistry and Molecular & Cellular Biology and Senior Associate Dean, Faculty and Academic Affairs, Georgetown University, for providing funds to Naha for data collection, from NSF grant DBI-0845523.

Author contributions: Vasudevan conceived the idea of this paper, carried out the analyses, and guided Naha, who was a graduate student at Georgetown University, to collect data on the sequences. Gana conceived the idea of estimating the likelihood of glycosylation using the classical regression framework. Vasudevan and Gana wrote the manuscript. Mazumder provided, using the SFAT tool, a data extract of potentially non-N-glycosylated sequences. Subsequently, a sample of Mazumder's data extract was cleaned, validated, and updated with structural information by Vasudevan and Naha for modeling. Goldman provided partial funding for the initial data collection phase of this project.

**Ridge Regression Estimated Linear Probability Model Predictions of N-glycosylation in Proteins with Structural and Sequence Data**

**Introduction**

N-linked glycosylation is a complex co– and post– translational modification in proteins that occurs most commonly in a defined Aspagarine-x-Serine/Threonine (NxS/T) sequon, where x is any amino acid other than Proline. Hereafter, the NxS/T sequon is referred to as the N-site. The N-site is commonly observed in proteins, some of which are known to be glycoproteins, and others are not. Glycosylation is a process in which sugars (simple and complex) get covalently linked via an *N*-Acetylglucosamine (*N*-acetyl-D-glucosamine, or GlcNAc, or NAG) to an Asn (N) residue forming an N-glycosidic bond (Ueda, 2013; Welti, 2013; Zhang, Yin, & Lu, 2012). However, non-canonical glycosylation sites (i.e., NxY, where Y is any amino acid) have also been observed in proteins (Chi, Y.H. et al., 2010; Crispin, M. et al., 2007; White, C.L. et al., 1995).

Glycans are produced from a small set of monosaccharide building blocks using a secondary metabolism, rather than being encoded in the genome itself. The glycans are built in a stepwise fashion by multiple enzymes and, thus, by multiple genes. The enzymatic machinery responsible for building and remodeling glycans, contributes to the complexity of glycan structures based on different patterns and linkages. Each glycosylation site within a population of glycoprotein molecules is typically occupied by a few major and additional minor glycan forms. This results in a diversity of glycan structures, and results in a single protein substrate with different glycans. This is referred to as the glycoprotein's *microheterogeneity*. These differences in protein glycosylation can have distinct biological effects (Chandler & Goldman, 2013; Pompach, P. et al., 2013). In addition, some glycoprotein molecules are glycosylated, while others are not. The percentage of a glycosylation site occupied by glycans within a distinct population of glycoprotein molecules can vary and affect the quantity of glycoforms observed in a particular context. This is referred to as the *occupancy* of a glycosylation site. The cause for this heterogeneity and varied occupancies is not well understood, and remain key questions in the glycomics field.

The importance of glycosylation is becoming widely recognized through studies aimed at understanding its role in several major diseases, specifically cancer (Breier, Gibalova, Seres, Barancik, & Sulova, 2013; Chang & Hung, 2012; Compte, Nunez-Prado, Sanz, & Alvarez-Vallina, 2013; Gruszewska & Chrostek, 2013; Peiris, D. et al., 2017; Fukami, K. et al., 2017; Cui, J. et al., 2018). This is not surprising, because glycosylation is a stepwise procedure of covalent attachment of oligosaccharide chains to proteins or lipids, and alterations in this process have been associated with malignant transformation (Goldman, R. et al., 2009; Potapenko, I.O. et al., 2010). Given that more than 50% of eukaryotic proteins are glycosylated (Apweiler, Hermjakob, & Sharon, 1999), and that glycosylation is important for protein stability, glycomics research is gaining visibility (Mochizuki, K. et al., 2007; Price, Powers, Powers, & Kelly, 2011; Zheng, Bantog, & Bayer, 2011; Ahmad, Moinuddin, Khan, & Ali, 2012; Lu, Yang, & Liu, 2012; Zou, Huang, Kaleem, & Li, 2013). This visibility is also evidenced by several resources aimed at providing glycan information like the GLYCOSCIENCEs.de web portal (Lutteke, T. et al., 2006), UniCarbKB (Campbell, M.P. et al., 2011), and the Glycome-DB (Ranzinger, Frank, von der Lieth, & Herget, 2009). In addition, considering that the glycosylation sequon is a common occurrence in proteins, a reliable statistical model combining sequence and structural attributes to predict which proteins are glycosylatable is essential for guiding experimental work. In this paper, a classical regression model is built to assess the likelihood of N-glycosylation based on sequence and structural attributes.

Key advances in the field of glycomics research also indicate that three dimensional structures play a role in stabilizing the N-sites (Hurtado-Guerrero & Davies, 2012; Nagae & Yamaguchi, 2012). Furthermore, there is evidence that disease causing mutations affect the stability of the protein's three dimensional structure, and, hence, its function (Busquets, C. et al., 2000; Musco, G. et al., 2000; Kim & Hahn, 2015; Taniguchi, T. et al., 2017). Recent advances in crystallization and structure solution methodologies have enabled one to overcome the technical difficulties involved in crystallizing protein-glycan complexes. A body of structural data reported in complex with glycans in the Protein Data Bank (PDB) make structural studies possible, and enable studies geared toward answering questions that are important to protein-glycan interactions. However, while there are several resources aimed at providing glycan information, a comprehensive collection, and detailed analysis, of all N-linked glycoproteins with structural information is still lacking.

The last decade has seen significant technological advancements in producing structures of glycoproteins with complex sugars. Statistical analysis on X-ray crystallographic structures have been carried out. These were aimed at identifying energetically favorable conformations for individual sugar linkages. This information is very important for understanding structural attributes of glycoproteins, because several of them are emanating as drug targets (Das, Biswas, & Khera, 2013; Yoo, J. et al., 2018). However, a detailed structural analysis on all available glycoprotein structures is not available. This present study is aimed at providing an in-depth structural and sequence analysis of N-glycosylated sites.

In this paper, based on the distribution of amino acids in the neighborhood of the N-site, and the structural attributes of the N-site that include Accessible Surface Area (ASA), secondary structural elements, main-chain phi-psi, turn types, location of the N-site in the primary sequence, and the nature of the glycan bound, the likelihood of N-glycosylation is predicted. A robust prediction methodology to assess the relative probability of N-glycosylation is essential for guiding experimental work.

**The Collected Data**

Beginning in the year 2014, 8,962 sequences with structural information were collected. These sequences are documented in an Excel spreadsheet: *glycos_public.xlsx* (available upon request). Of the 8,962 sequences, 5,054 are human; and of these human sequences, 2,422 are known to be N-glycosylated[2].

---

[2] Of the 2,422 sequences, 222 are also fucosylated; and of the 3,906 sequences, 261 are fucosylated. We define structures as fucosylated if a fucose is attached to the first NAG that is covalently attached to the N-site. The introduction of bisecting GlcNAc and core fucosylation in N-glycans is essential for functional regulation of glycoproteins. Core fucosylation, which is a very important glycan modification observed in cancer and other diseases, can be modeled as we model N-glycosylation, but that modeling is planned for later. Fucosylated glycans are increasingly drawing attention as biomarkers because of their dramatic increase in expression in both cancers and inflammation (Moriwaki & Miyoshi, 2010; Liu, L. et al., 2013; Nishima, Miyashita, Yamaguchi, Sugita, & Re, 2012; Isaji, T. et al., 2010; Ferrantelli, E. et al., 2018).

**Ridge Regression Estimated Linear Probability Model Predictions of N-glycosylation in Proteins with Structural and Sequence Data**

*Binding set*: This set consists of structures that have at least one *N*-acetylglucosamine (*N*-acetyl-D-glucosamine, or GlcNAc, or NAG) covalently linked to the asparagine residue of the N-site. These sequences have the canonical NxS/T sequon. Structures with a non-canonical N-site or without a mapped UniprotKB accession in the UniprotKB database, and sequences which did not have flanking +10 or -10 amino acids on either side of the N-site are excluded. Structural data was obtained from the Protein Data Bank (PDB) (Berman, H.M. et al., 2000), and the PDB-ID codes used are provided in *glycos_public.xlsx* in the column labeled *PDBID*. The sequence information and taxonomy information for the data used in the analysis were extracted from the UniprotKB database (www.uniprot.org) and shown in columns labeled *Uniprot_Accession_No* and *Protein_Organism* in *glycos_public.xlsx* (Magrane & UniProt Consortium, 2011).

*Human Set and Non-human Set:* Sequences that have *Protein_Organism* in *glycos_public.xlsx* as Homo Sapiens constitute the human set and the remaining the non-human set.

*Non-binding set*: This is the set of sequences that contain the N-site but are, thus far, not experimentally deemed to be glycoproteins.

*ASA*: The solvent accessibilities for the Asn residue in the N-site are obtained using the ASAView tool and database (see column labeled *ASA* in *glycos_public.xlsx*) (Ahmad, Gromiha, Fawareh, & Sarai, 2004).

*Ligand sequence*: The ligand information was obtained from the PDBsum database (see column labeled *Ligand_Sequence* in *glycos_public.xlsx*) (Laskowski, R.A. et al., 1997).

*Turn types*: The ProMotif database was used to obtain the turn types. We followed the definitions described in: http://www.ebi.ac.uk/thornton-srv/databases/cgi-bin/pdbsum/GetPage.pl?pdbcode=n/a&template=doc_promotif.html. These are in *glycos_public.xlsx* under the column labeled *Turn_Type*. Based on the definitions in ProMotif, 24 different secondary structural elements are defined. As per the definition of secondary structure that the ProMotif program uses, helices and strands are computed from main-chain hydrogen bonding patterns. Beta and gamma turns are defined by phi-psi torsion angle combinations, except where all of the residues concerned have already been assigned to a helix. Several pairings based on the occurrence of an amino acid sharing itself with different structural elements are defined. For example in PDB-ID 1A0H, site 373 (chain B) is listed as residue in both the BETA_HAIRPIN section and the strand section of ProMotif. Hence, this is defined as BETA_HAIRPIN_STRAND.

*Amino acid classification used for frequency distribution*: The definitions in the ASAView tool are used to classify the amino acids into five categories: positively charged residues (R, K, H), negatively charged residues (D, E), polar uncharged residues (G, N, Y, Q, S, T, W), Cysteine (C), and hydrophobic residues (L, V, I, A, F, M, P).

*Out-of-sample sequences*: A sample of 152 not yet curated proteins is held out as the out-of-sample data. These are labeled "yes" in column *glycosed_later* in *glycos_public.xlsx*.

Page **4** of **20**

**Ridge Regression Estimated Linear Probability Model Predictions of N-glycosylation in Proteins with Structural and Sequence Data**

## Some Properties of the Collected Data

The counts of sequences and the unique proteins they represent are shown in Table 1.

Table 1: Counts of sequences and unique proteins in the collected data (i.e., *glycos_public.xlsx*)

| Data | N-site count | Unique protein count* |
|---|---|---|
| All data (N-glycosylated and non-N-glycosylated) | 8,962 | 2,216 |
| Only non-humans (all of these are N-glycosylated) | 3,908** | 473 |
| Only humans (N-glycosylated) | 2,422 | 275 |
| Only humans that are considered N-glycosylated, but have only a single-sugar-bound | 1,339 | 211 |
| Only humans that are considered N-glycosylated, but have more than one sugar-bound | 1,083 | 178 |
| Out-of-sample not yet curated proteins (considered not N-glycosylated, by default) | 154** | 99 |

* Sequences with unique Uniprot Accession Numbers.  ** Two of these have missing N-glycosylated statuses and are ignored.

*Distribution of amino acids around the N-site for glycosylated human sequences:*

Earlier studies, based on small datasets, have identified position-specific amino acid preferences around the N-site. For example, an increased occurrence of aromatic residues immediately before the N-site was observed, but this probability of occurrence decreased in positions immediately following the N-site (Petrescu, Milac, Petrescu, Dwek, & Wormald, 2004). In order to see if earlier observations still hold, the counts of amino acids flanking +10 and –10 on either side of the N-site were obtained using *glycos_public.xlsx*. Before counting, N-glycosylated sequences having exactly one-sugar bound were removed. This produced a subset of 1,083 human sequences. Percentage counts are shown in Table 2. In each position (except +2, of course) flanking the N-site, about 70% or more of the amino acids are polar uncharged or hydrophobic. These results agree with previous studies regarding the preference of hydrophobic amino acids near the N-site. However, in addition, a preference for polar uncharged amino acids near the N-site is also found in this present study.

**Ridge Regression Estimated Linear Probability Model Predictions of N-glycosylation in Proteins with Structural and Sequence Data**

Table 2: Percentage distribution of aromatic residues around the glycosylated N-site

| Amino Acid Position | Positive | Negative | Polar | Cystein | Hydrophobic |
|---|---|---|---|---|---|
| −10 | 12.56 | 9.88 | 33.89 | 4.52 | 39.15 |
| −9 | 11.73 | 8.31 | 38.50 | 2.31 | 39.15 |
| −8 | 5.45 | 12.83 | 40.81 | 1.02 | 39.89 |
| −7 | 9.97 | 10.06 | 36.47 | 1.39 | 42.11 |
| −6 | 8.68 | 11.54 | 33.70 | 1.02 | 45.06 |
| −5 | 11.17 | 7.85 | 31.58 | 1.20 | 48.20 |
| −4 | 13.20 | 9.60 | 41.92 | 1.48 | 33.80 |
| −3 | 16.53 | 13.48 | 28.62 | 2.31 | 39.06 |
| −2 | 9.88 | 3.97 | 34.35 | 0.83 | 50.97 |
| −1 | 17.91 | 7.39 | 30.47 | 3.60 | 40.63 |
| 0 (N-site) | | | | | |
| +1 | 12.28 | 10.25 | 34.53 | 2.68 | 40.26 |
| +2 | 0 | 0 | 100 | 0 | 0 |
| +3 | 8.03 | 16.71 | 33.89 | 3.42 | 37.95 |
| +4 | 8.22 | 10.71 | 29.18 | 2.22 | 49.68 |
| +5 | 7.66 | 13.30 | 26.69 | 2.77 | 49.58 |
| +6 | 16.07 | 9.51 | 32.23 | 0.65 | 41.55 |
| +7 | 13.39 | 7.48 | 34.07 | 3.97 | 41.09 |
| +8 | 8.59 | 16.81 | 30.38 | 4.16 | 40.07 |
| +9 | 9.97 | 10.71 | 39.52 | 2.86 | 36.93 |
| +10 | 10.71 | 13.94 | 34.35 | 3.42 | 37.58 |

*Secondary structural elements of the glycosylated N-site for humans:*

It has been reported in the literature that the N-site is, in general, predominantly found to occur as part of loop regions and to a lesser extent in helices and strands. This is not surprising given the preference of ASN to be in loops (Liau, Sleight, Pitha, & Peroutka, 1991). However, the present analysis of glycosylated human sites showed that about 23% of the N-sites are found in loops, 11% in helices, 24% in strands, and 41% in beta turns. In the literature, the likelihood of sugars binding to the N-site has not been differentiated by turn types; however, in this paper, turn types play a significant role in this regard.

*Glycan conformations and fold types*:

The conformation of the protein-glycan linkage, as measured by phi-psi of the N (Asn) of the N-site, has been shown to be one of the determining factors differentiating bound vs. unbound sugars. A plot of the phi-psi of N falls into four distinct clusters with a few outliers (Figure 1). In addition to the phi-psi, it is found that the glycan conformations vary among the different structures (data not presented). The glycoprotein structures analyzed in this study belong to about 30 different structural fold types as per SCOP and Superfamily classifications (Murzin, Brenner, Hubbard, & Chothia, 1995). However, in order to demonstrate the role of conformational flexibility of N-glycans in protein–glycan interactions, further in-depth analysis is needed to correlate this with the functions of these glycoproteins and fold-types. A ligand-centric approach for analyzing ligand conformations has been developed (Gana, Rao, Huang, Wu, & Vasudevan, 2013). A comprehensive analysis of all glycan conformations by applying the ligand-centric approach may be a topic of interest for future research.

**Ridge Regression Estimated Linear Probability Model Predictions of N-glycosylation in Proteins with Structural and Sequence Data**

Figure 1: Phi-Psi plot for the human N-glycosylated sequences

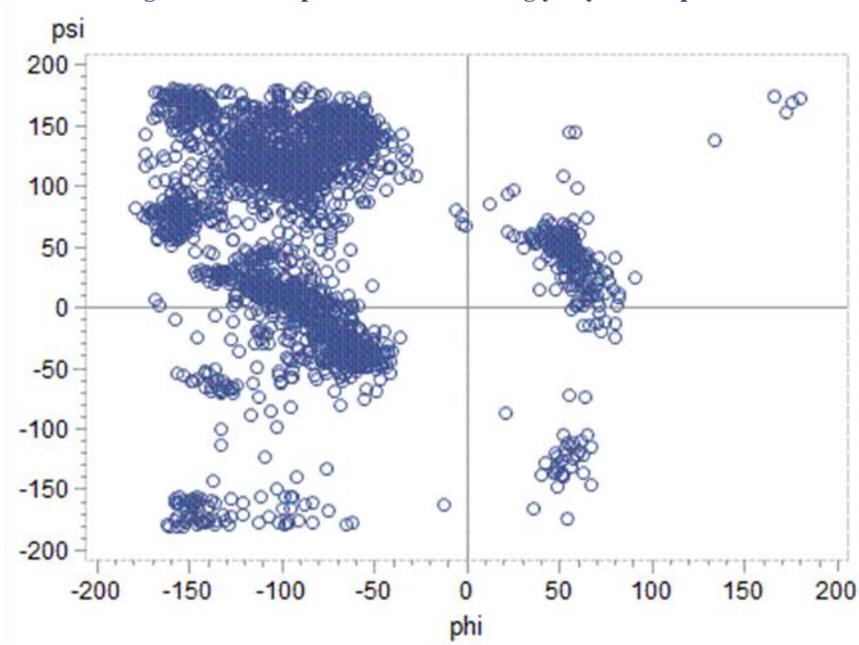

## Linear Probability Modeling to Predict N-glycosylation

In the absence of experimental evidence, little is known about how to predict whether or not N-linked glycosylation, occurs in proteins. Although it is well-known that the NxS/T sequon is, generally, a necessary (but, not sufficient) condition for N-glycosylation, little is known about how the distribution of amino acids in other positions near the N-site influence the likelihood of N-glycosylation.

This paper assumes that the occurrence of specific amino acids in specific positions around the neighborhood of amino acid N (the "center", satisfying the known canonical N-site) contribute to determining whether or not a sequence is sugar linked.

Based on the collected data (*glycos_public.xlsx*), the neighborhood of N around the NxT site is defined to be 10 positions to the left, and 10 to the right, of N. Longer neighborhoods of N were not considered. The $i^{th}$ position to the left of N (N-terminal side) is named *minus$_i$*. The $i^{th}$ position to the right of N (C-terminal side), is named *plus$_i$*. Data is only selected for which *plus$_2$* equals S or T, because it is well-known that the presence of the motif NxS/T is a necessary condition for having a sugar linked sequence. If amino acid $\alpha$ (say) occurs in the $i^{th}$ position, then an indicator (or binary) variable that takes the value 1 or 0 is defined as follows: *minus$_{i\alpha}$* = 1 if $\alpha$ is to the left of N and *plus$_{i\alpha}$* = 1 if $\alpha$ is to the right of N; otherwise, these indicator variables take the value 0. Thus, 200 (20 amino acids × 10 positions) indicator variables define the left or right neighborhood of N; *plus$_{2T}$* = 1 if amino acid T is in *plus$_2$*, and 0 otherwise. The additional assumption made is that the position of N relative to the total length of the sequence is also a significant determinant of whether or not the sequence is sugar linked. This assumption is quantified by defining, for sequence *j*, the variable *pos$_j$* as the ratio of the position of N to the length of the sequence; thus, *pos$_j$* is a number between 0 and 1, with numbers close to 0 indicating that N is closer to the beginning of the sequence than it is to the end of it. Finally, it is assumed that the structural

**Ridge Regression Estimated Linear Probability Model Predictions of N-glycosylation in Proteins with Structural and Sequence Data**

information (ASA values, turn types, and phi and psi angles) on the sequences are additional explanatory variables contributing toward N-linked glycosylation.

If $Y_j$ is another indicator variable (i.e., the binary dependent variable) that takes the value 1 if sequence $j$ is sugar linked and 0 otherwise, then the postulated regression model can be specified as follows:

$$Y_j = \beta_0 + \Sigma_\alpha \Sigma_i \beta_{i\alpha} \, minus_{i\alpha} + \Sigma_\alpha \varphi_{1\alpha} \, plus_{1\alpha} + \varphi_{2T} \, plus_{2T} + \Sigma_\alpha \Sigma_i \varphi_{i\alpha} \, plus_{i\alpha} + \lambda \, pos_j$$
$$+ \delta \, ASA_j + \pi \, (ASA\_zero_j) + \Sigma_k \theta_k \, (Turn \, Type_k) + \omega_1 \, (psi \, angle) + \omega_2 \, (phi \, angle)$$

where $\beta_0$, $\beta_{i\alpha}$, $\varphi_{i\alpha}$, $\lambda$, $\delta$, $\pi$, $\theta_k$, $\omega_1$ and $\omega_2$ are constants (coefficients) estimated using the data; $\Sigma_\alpha$, $\Sigma_i$ denote summations over amino acid and position combinations, respectively, and $\Sigma_k$ denotes summation over turn types, that are found to be statistically significant (i.e., do not have a coefficient of zero in the model). The variable *ASA_zero* is a dummy variable that takes the value 1 if ASA value is zero, and 0 otherwise – this dummy variable is essential to capture the additional decrease in the likelihood of N-glycosylation if the sequence is not exposed at all (i.e., *ASA = 0*)[3]. The variables on the right hand side of the above equation are the explanatory variables that are assumed generate the outcome (i.e., $Y_j$). This regression, a linear probability[4] model[5] (LPM), estimates the probability (Goldberger, 1964) that a sequence is sugar linked given the explanatory variables.

The LPM is estimated using the least squares (LS) method, which produces consistent estimates of the coefficients (McGillivray, 1970). Although LPM predictions may sometimes be negative or greater than unity, this is not an asymptotic problem (Amemiya, 1977); and given the purpose of this paper, which is to rank order sequences in terms of sugar linked probabilities, negative LPM predictions are interpreted as nearly 0 (i.e., the sequence is not sugar linked) and LPM predictions greater than unity are interpreted as nearly 1 (i.e., the sequence is sugar linked)[6].

---

[3] Alternatively, one can drop *ASA_zero* as an explanatory variable, and, thereby, allow its effect to be distributed among the other coefficients; but we did not do so, because it is of interest to see its marginal contribution toward discouraging N-glycosylation.

[4] Our use of the complex concept called "probability" (Ergodos, 2014) in this paper does not imply that we are saying there is a fundamental or intrinsic randomness in how a sequence folds and becomes sugar linked; rather, we are saying that given the underlying complexity of this process, our methodology, which depends on probabilistic ideas, may represent a reasonable approximation (via rank-ordering) to facilitate predicting whether a sequence is N-glycosylated. Another way of saying this is that when a "fair" coin (Keller, 1986) is tossed the outcome (heads or tails) is exact, but given the mathematical complexity of, and the instruments needed for, making this prediction based on the laws of physics (e.g., Newton's Laws), we use the language of probability, instead of physics, to estimate the outcome (Diaconis & Mazur, 2003). In particular, we say the probability of heads is 50%. But, although this statement is false in theory, it is an approximation that can be quite useful in practice, and for guiding experimental work.

[5] The LPM is $Y = Xb + u$, but where $Y$ only takes the values 0 and 1. If $E(u) = 0$, then each $u_i$ has variance $E(Y_i)(1 - E(Y_i))$. In 1964, Goldberger suggested estimating $E(Y_i)$ by ordinary least squares (OLS), and then re-estimating the model by weighted least squares (WLS) to achieve homoscedasticity (i.e., constant residual variance).

[6] One method to guarantee the LPM predictions lie between 0 and 1 is to estimate the LPM using the LS method with inequality restrictions (Judge & Takayama, 1966): for example, minimizing the sum of squared errors (or another objective function like mean absolute error) subject to the constraints $0 \leq Xb \leq 1$. But this approach was not pursued in this study, because the sampling properties of the resultant estimator are not available. An advantage of using the LPM is that it allows us to estimate coefficients for an amino acid and position combination that may be perfectly correlated with one value of $Y_j$ – for example, if such a coefficient exists (e.g., for a motif only present in sugar linked sequences), it cannot be estimated in logit models (Caudill, 1988).

**Ridge Regression Estimated Linear Probability Model Predictions of N-glycosylation in Proteins with Structural and Sequence Data**

Because little is known about how amino acid occurrences by positions in the neighborhood of N influence the probability of a sequence becoming sugar linked, significant (at the 5% level) explanatory variables in the LPM are selected using the stepwise selection method (Efroymson, 1960) [7] in conjunction with 10-fold cross-validation[8].

Because the LS estimated LPM produces heteroscedastic errors (i.e., residuals with varying variances), it was confirmed that stepwise selected significant variables continued to remain significant when re-evaluated in terms of heteroscedasticity-consistent standard errors (White, 1980), as a first approximation. Because the exact functional form of the LPM error variance is known, ridge regression[9] (RR) is applied to the stepwise selected LPM to reconfirm that all variables continue to be statistically significant. This is done by re-estimating the stepwise selected LPM by RR so the predicted values of $Y_j$ are forced to lie in the 0-1 interval (Gana, 1995)[10]. Checks for outliers[11] were conducted, because it is important that "unusual" sequences not unduly influence the rank ordering of other sequences in terms of their probability of becoming sugar linked. RR estimation of the LPM has been found to be competitive with other LPM estimation methods and logit[12] regression (Monyak, 1998)[13].

---

[7] The stepwise procedure starts with no variables and adds them one at a time according to their partial *F*-statistics (Hocking, 1976) until either all variables are included or no excluded variable's partial *F*-statistic is significant. Although the stepwise method, per se, cannot guarantee finding optimal subsets of significant variables (Hoerl, Schuenemeyer, & Hoerl, 1986), we used it in conjunction with 10-fold cross-validation to minimize the prediction residual sum of squared errors (PRESS), which has a point of contact with the Brier Score (Brier, 1950; Murphy, 1973).

[8] In cross-validation (Wallace & Mosteller, 1963; Tukey & Mosteller, 1968) the data is partitioned into two subsets. One subset (a.k.a. the "training" data) is used to estimate ("train") the model; and the other subset (a.k.a. the "validation" data) is used to measure ("validate") the accuracy of model predictions using a chosen metric (e.g., PRESS). The subsets must be selected in a manner such that each data point has a chance of being validated against. In 10-fold cross-validation, the data is partitioned, randomly, into 10 equal (or nearly equal) subsets (a.k.a. "folds"). Then model training and validation are performed 10 times such that each time a different fold of the data is used for validation while the remaining 9 folds are used for training. In our case, 10 samples of PRESS would be available after the cross-validation exercise. The chosen LPM is the one that yields the minimum average PRESS. The idea of cross-validation is old (Larson, 1931). Its use for model selection, rather than just for model validation, is, relatively, more recent (Stone, 1974; Geisser, 1975).

[9] The RR (Hoerl & Kennard, 1970; Hoerl & Kennard, 1990; Qannari, Vigneau, & Semenou, 1990; Gruber, 2010) estimator of *b*, $b_R$, is given by $(X^TX + kI)^{-1}X^TY$, where $k \geq 0$, the RR tuning parameter, is the smallest constant for which all of the resultant LPM predictions, $X b_R$, lie between 0 and 1. The classical bisection method can be used to calculate such a value of $k$. Next, WLS is used to re-estimate *b* by using the weights: $X b_R ( 1 - X b_R )$. This WLS estimate of *b* is denoted by $b_W$. Because $X b_W$ is not necessarily constrained to lie in the range 0-1, weighted RR (WRR) can be used, instead of WLS, to re-estimate *b*. Let $b_{WR}$ denote the WRR estimate of *b*. Generally, RR is used to correct for multicollinearity. The usual RR estimator (i.e., for a continuous regressand), which employs optimal linear shrinkage (Frank & Friedman, 1993) to improve prediction properties, yields the same *t*-statistics and *F*-ratios as does the LS estimator (Obenchain, 1977).

[10] Other methods (Goldfeld & Quandt, 1972; Hensher & Johnson, 1981; Mullahy, 1990; Horrace & Oaxaca, 2006) to ensure there are no negative LPM residual variances were not pursued in this paper. One ad-hoc method simply sets LPM predictions greater than 1 to a number close to 1 (such as 0.999) and negative LPM predictions to a number close to 0 (such as 0.001). Another ad-hoc method uses the absolute values of the OLS estimated residual variances to do the WLS estimation. Goldfeld and Quandt (1972) proposed only using those observations having OLS estimates between 0 and 1 to do the WLS estimation. Hensher and Johnson (1981) proposed bounding the weights and assigning negative weights a constant value. Mullahy (1990) proposed a quasi-generalized least squares estimator which is a generalization of the Goldfeldt-Quandt and Hensher-Johnson estimators.

[11] Although little work on the impact of outliers on RR has been done (Walker & Birch, 1988; Chalton & Troskie, 1992), an important unpublished doctoral dissertation (Saccucci, 1985) indicates that RR is robust to the influence of outliers. Because this work of Michael Saccucci is of importance here, it is summarized. Saccucci considered the case of variance inflated outliers (VIOs) in the usual linear model where *Y* is a continuous regressand. A VIO is an observation whose residual variance is $\sigma^2 w$, where $w > 1$ is a constant. Saccucci assumed that given *n* observations, *m* of them are VIOs each with residual variance $\sigma^2 w$. He assumed that the remaining $n - m$ observations each have residual variance $\sigma^2$. Let $X_m$ denote the sub-matrix of *X* containing the VIOs. Saccucci showed that the mean square error (MSE) of the RR estimated *b* under this assumption of VIOs, is equal to the MSE of the RR estimated *b* under the assumption of no outliers plus $\sigma^2(w-1)$ times the sum of the diagonal elements of the following matrix: $( X^TX + kI )^{-1}X_m^TX_m( X^TX + kI )^{-1}$, where, as usual, *I* denotes the identity matrix, *T* denotes the transpose operator, and $k \geq 0$ is a constant. He showed that this matrix (which is the additional MSE for the RR estimator) decreases monotonically with *k*. He showed that there always exists (∃) a $k > 0$ for which the MSE of the RR estimated *b* under his assumption of VIOs, MSE ( $b_R$ | VIOs ), say, is less than the MSE of the least squares estimated *b*, $b_{LS}$, under his assumption of VIOs, MSE ( $b_{LS}$ | VIOs ), say. That is, he established that the following inequality holds: ∃ $k > 0$ ∋ MSE ( $b_R$ | VIOs ) < MSE ( $b_{LS}$ | VIOs ). This inequality is a generalization of the original existence theorem of Hoerl and Kennard (1970). Saccucci used simulation to show that his result holds (with probability > 0.5) for the

**Ridge Regression Estimated Linear Probability Model Predictions of N-glycosylation in Proteins with Structural and Sequence Data**

The RR predicted values of $Y_j$, which lie in the range 0-1, are used to estimate the LPM error variances: (RR predicted $Y_j$ ) × (1 – RR predicted $Y_j$ ). The optimal value of the RR tuning parameter, *k*, for which all of the aforesaid predictions fall in the range 0-1 is 3.6469. Using these variances as weights, the LPM is re-estimated using classical weighted LS (WLS). If variables with *p*-values greater than 5% were found, the entire LPM estimation process was redone. The selected LPM is the one for which all variables are statistically significant at the 5% level.

For the selected LPM, Cook's (Cook, 1977) distances were calculated to flag sequences present in the data, if any, that may unduly influence the selected LPM; such "influential" or "unusual" sequences may be present as a result of data errors (e.g., sugar linked sequences recorded as not sugar linked) or other reasons. Variance inflation factors (VIF) are computed to ensure that they are less than 10, which make the problem of multicollinearity a non-issue (Marquardt, 1970). For example, severe multicollinearity can occur if an amino acid in a particular position is determined by amino acids in other positions; or, if structural variables are strongly driven by amino acids and the positions they occupy; or combinations of these potential situations.

---

values of *k* selected using several proposed algorithms (Lawless & Wang, 1976; Hoerl, Kennard, & Baldwin, 1975; Hoerl & Kennard, 1976; Dempster, Schatzoff, & Wermuth, 1977).

[12] It should be noted that the logit model functional form guarantees [by defining $Pr(Y_j = 1) = e^{Xb} \div (1+e^{Xb})$] that the estimates of $Y_j$ lie in the 0-1 interval, but under maximum likelihood estimation it can be highly sensitive to outliers (Pregibon, 1981), and may require accounting for multicollinearity (Le Cessie & Van Houwelingen, 1992; Schaefer, Roi, & Wolfe, 1984). However, accounting for outliers and multicollinearity in logit models are more complex than doing so in linear models. Furthermore, the natural attractiveness of the logit model to constrain the estimates of $Y_j$ in the 0-1 interval does not necessarily imply that the logit specification is superior to the LPM or that the logit imposed nonlinearity on the data is fundamental to understanding the process generating the data.

[13] We can look at Saccucci's work in a new way and, thereby, connect it to the idea of the RR estimated LPM. Each residual, $u_i$, under the LPM has variance $x_ib(1 − x_ib)$. Hence, the LPM can be viewed as a linear model with observations having distinct variances, in Saccucci's sense. Therefore, one can conjecture that: ∃ *k* > 0 ∋ MSE ( $b_R$ | LPM ) < MSE ( $b_{LS}$ | LPM ), where MSE ( · | LPM ) denotes the MSE of "· " under the LPM assumption. Furthermore, if we can show that this result also holds for the optimal *k* (i.e., the smallest value of *k* > 0 for which all of the LPM predictions are between 0 and 1), then we have a stronger case for considering the RR estimated LPM useful. Now, it is also known that the RR estimator can also improve prediction properties (Theobald, 1974). Thus, it would be good to know whether the following holds: ∃ *k* > 0 ∋ MSE ( $Xb_R$ | LPM ) < MSE ( $Xb_{LS}$ | LPM ). Again, this result will be stronger if we can also show that it holds for the optimal value of *k*.

In an important unpublished doctoral dissertation, John Monyak (1998) showed that the abovementioned conjectures hold true. Monyak showed that the RR estimated LPM is consistent. His simulation results indicate that the RR estimated LPM is superior to the least squares estimated LPM both in terms of coefficient and prediction MSEs. In particular, at the optimal value of *k*, Monyak's key simulation results included: MSE ( $b_R$ | LPM ) and MSE ( $b_{WR}$ | LPM ) are 77.5% (standard error = 3.6%) and 80.2% (standard error = 3.9%) of MSE ( $b_{LS}$ | LPM ), respectively; MSE ( $Xb_R$ | LPM ) and MSE ( $Xb_{WR}$ | LPM ) are 87.5% (standard error = 1.8%) and 87.8% (standard error = 2.0%) of MSE ( $Xb_{LS}$ | LPM ), respectively; MSE ( $b_R$ | LPM ) < MSE ( $b_{LS}$ | LPM ), 60.5% of times; and MSE ( $Xb_R$ | LPM ) < MSE ( $Xb_{LS}$ | LPM ), 66.0% of times.

When the RR estimated LPM is compared with some of the other proposed LPM estimators (like the ad-hoc, Goldfeld & Quandt, and Mullahy estimators), Monyak's simulation results indicate that its superiority, in terms of MSEs, continues to hold. The Goldfeld-Quandt and Mullahy methods produce coefficient and prediction vector MSEs that are about 118% each, of the corresponding least squares MSE values. The ad hoc method (i.e., rounding predictions to 0.999 or 0.001) produces coefficient and prediction MSEs of 98.3% and 98.6% of the corresponding least squares MSE values, respectively. In Monyak's simulation, least squares produces predictions outside the range 0-1, 42% of times.

**Ridge Regression Estimated Linear Probability Model Predictions of N-glycosylation in Proteins with Structural and Sequence Data**

The above described variable selection process used to select a LPM constitutes one model search attempt, because for each attempt, 10-fold cross-validation is done randomly with a particular "seed" value. To understand the variability in variable selection, 20 different seed values were used – that is, 20 model search attempts were made. Because little is known about the process generating N-linked glycosylation, setting the seed to a constant value in the interest of keeping the selected explanatory variables the same was not considered appropriate. Over the 20 model search attempts, the percentage of significant (at the 5% level) explanatory variables entering potential models less than 100% of the time are flagged in Figure 2.

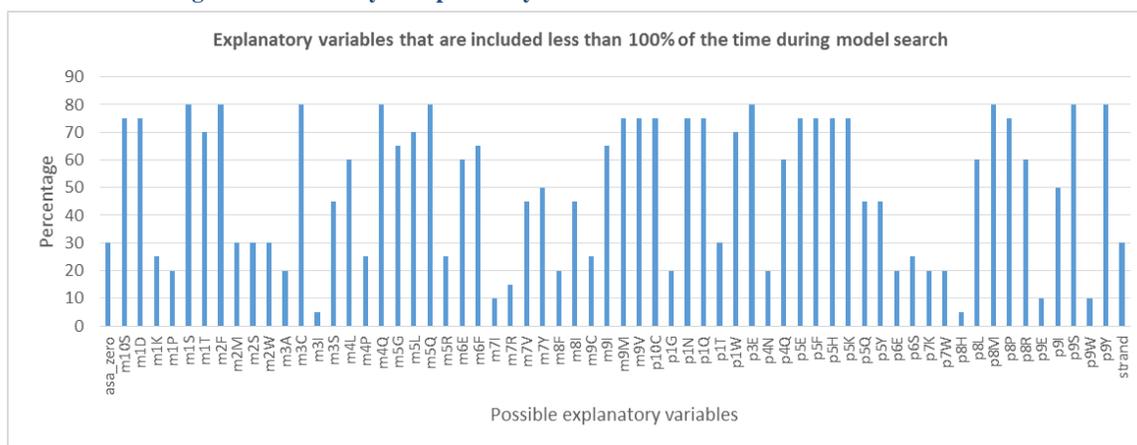

Figure 2: Variability in explanatory variables selected for inclusion in the LPM*

* *miα* and *piα* are abbreviations for *minusiα* and *plusiα*, respectively; where *α* is an amino acid and *i* its position

The selected LPM, shown in Table 3, is estimated on 1,083 (human) sugar linked sites and 2,480 non-sugar-linked sites. The sugar-linked sites have more than one sugar bound. N-glycosylated sites with only a single sugar are deleted from the estimation data because these may be false positives given the known artifacts from crystallization. The number of amino acid and position combinations found to be significant in the selected LPM is 117.

The selected LPM yields an in-sample Kolmogorov-Smirnov (K-S) statistic[14] (Smirnov, 1939; Smirnov, 1948; Kolmogorov, 1933; Feller, 1948; Doob, 1949) value of about 74.3%, which indicates that the model is not an unreasonable one. The K-S statistic measures the maximum separation between the distributions of the estimated $Y_j$ for $Y_j = 1$ and $Y_j = 0$, respectively – the greater the separation, the better the model – for this LPM the maximum separation occurs at a predicted probability value of about 40%. The average (in-sample) predicted probabilities of N-glycosylation when $Y_j = 0$ and $Y_j = 1$ are about 11% and 71%, respectively. The standard deviations of predicted probabilities of N-glycosylation when $Y_j = 0$ and $Y_j = 1$ are about 23% and 27%, respectively.

---

[14] Such statistics are popular in the credit scoring literature where how well a model predicts a binary outcome is of critical importance (Hand & Henley, 1997; Thomas, 2000; Crook, Edelman, & Thomas, 2007). An important alternative to the K-S statistic is the Gini coefficient (Gini, 1912; Gini, 1921; Ceriani & Verme, 2012), which has a point of contact with Somers' *D* statistic (Somers, 1962; Nelsen, 1998), and the *C*-statistic [(1+ Gini) ÷ 2] popular in the medical literature (Austin & Steyerberg, 2012; Uno, Cai, Pencina, D'Agostino, & Wei, 2011). The in-sample Gini coefficient and *C*-statistic (a.k.a. AUC) for our LPM is about 89.2% and 95%, respectively. The *C*-statistic represents the probability that a randomly selected N-glycosylated sequence has a higher probability of being N-glycosylated than does a randomly selected non-N-glycosylated sequence.

**Ridge Regression Estimated Linear Probability Model Predictions of N-glycosylation in Proteins with Structural and Sequence Data**

The Durbin-Watson (DW) statistic (Durbin & Watson, 1950; Durbin & Watson, 1951) for the LPM was a bit low (about 1.1, rather than between 1.5 and 2). This may indicate the existence of missing explanatory variables (Gujarati, 1995), and, thus, room for model improvement or statistical fine-tuning. Another possible implication of the relatively low DW value is that there is some "time-series" property embedded in the creation of the sequences, which the model has ignored. That is, there may be intra-positional amino acid interactions: an amino acid in a particular position may influence or interact with another amino acid in another position (within the same protein).

Table 3: The selected LPM for predicting the probability of N-linked glycosylation

| Variable* | $\beta$ | $|t|$ | Variable* | $\beta$ | $|t|$ | Variable* | $\beta$ | $|t|$ | Variable* | $\beta$ | $|t|$ |
|---|---|---|---|---|---|---|---|---|---|---|---|
| intercept | 0.0628 | 2.36 | m6F | 0.0656 | 2.3 | p3E | 0.0523 | 2.55 | p9A | -0.1180 | 4.09 |
| m1C | 0.1081 | 3.02 | m6H | 0.1284 | 3.51 | p3K | -0.0829 | 3.19 | p9I | -0.0556 | 2.11 |
| m1D | -0.0991 | 3.38 | m6P | 0.2246 | 9.39 | p3P | -0.1993 | 5.55 | p9N | 0.0947 | 3.37 |
| m1E | -0.0842 | 3.42 | m6S | 0.0883 | 3.94 | p3Q | 0.0938 | 3.64 | p9P | 0.0744 | 2.87 |
| m1I | -0.1335 | 5.28 | m6T | 0.0922 | 3.41 | p3Y | 0.1243 | 4.4 | p9S | 0.0469 | 2.13 |
| m1N | -0.1664 | 4.2 | m6W | 0.3372 | 8.54 | p4I | 0.1664 | 7.86 | p9T | 0.1224 | 5.01 |
| m1S | -0.0699 | 2.54 | m7A | 0.1228 | 5 | p4K | -0.0851 | 3.02 | p9Y | -0.1106 | 3.71 |
| m1T | -0.0625 | 2.45 | m7E | -0.0996 | 4.07 | p4L | 0.0698 | 3.8 | p10A | -0.1216 | 5.04 |
| m2A | 0.1334 | 5.27 | m7G | 0.0911 | 3.55 | p4Q | 0.0609 | 2.2 | p10C | 0.1049 | 2.67 |
| m2D | -0.1370 | 4.76 | m7K | -0.0919 | 3.69 | p4R | -0.0727 | 2.8 | p10F | 0.1188 | 4.91 |
| m2E | -0.1095 | 3.7 | m7N | 0.1344 | 4.88 | p4T | 0.1106 | 4.22 | p10L | -0.0591 | 2.74 |
| m2F | 0.0749 | 3.06 | m7S | 0.0654 | 2.82 | p4V | -0.0997 | 4.29 | p10Q | 0.0935 | 3.14 |
| m2G | -0.1050 | 3.89 | m7T | 0.0964 | 3.61 | p5A | 0.0686 | 2.95 | p10R | -0.0799 | 2.99 |
| m2Q | 0.1385 | 5.08 | m7V | -0.0474 | 2.09 | p5C | 0.1748 | 4.53 | pos | -0.2190 | 10.02 |
| m2V | 0.0907 | 4.44 | m7Y | -0.0876 | 2.55 | p5E | -0.0689 | 2.95 | ASA | 0.0015 | 6.64 |
| m2Y | 0.1736 | 6.57 | m8I | -0.0565 | 2.19 | p5F | -0.0969 | 3.41 | ASA_zero | -0.1180 | 2.75 |
| m3C | 0.1389 | 3.31 | m8K | -0.0745 | 2.95 | p5H | -0.1187 | 3.16 | IV | 0.0798 | 4.44 |
| m3D | -0.0777 | 2.86 | m8T | 0.0750 | 3.2 | p5K | -0.0732 | 2.68 | VIII | -0.1515 | 3.73 |
| m3H | 0.0822 | 2.49 | m8Y | 0.0975 | 3.7 | p5Q | -0.0631 | 2.16 | Strand | 0.0636 | 2.3 |
| m3S | 0.0524 | 2.2 | m9I | -0.0743 | 3.14 | p5V | 0.0497 | 2.17 | Beta Bulges | -0.2576 | 3.05 |
| m3W | 0.1863 | 5.24 | m9M | -0.1424 | 2.81 | p5Y | -0.0888 | 2.31 | BH** | -0.2103 | 6.9 |
| m4G | -0.0642 | 2.59 | m9S | 0.0906 | 3.9 | p6H | 0.1456 | 4.82 | BH Strand | 0.6776 | 26.54 |
| m4I | -0.0880 | 3.09 | m9V | -0.0707 | 2.88 | p6I | 0.1188 | 5.58 | BH Loop | -0.1854 | 4.11 |
| m4L | 0.0480 | 2.47 | m10N | 0.0801 | 3.14 | p6K | -0.0824 | 3.07 | Helix BH | 0.7658 | 10.52 |
| m4Q | -0.1086 | 3.67 | m10R | 0.0716 | 2.76 | p6M | -0.1504 | 3.95 | phi | -0.0003 | 3.24 |
| m4S | 0.0914 | 4.28 | m10S | 0.0548 | 2.41 | p6W | 0.1346 | 3.36 | psi | 0.0002 | 2.37 |
| m4V | -0.1123 | 4.51 | m10V | 0.0691 | 3.3 | p7C | 0.1918 | 5.41 | | | |
| m5F | 0.0998 | 3.51 | m10W | 0.1238 | 3.11 | p7P | 0.1171 | 4.58 | | | |
| m5G | 0.0637 | 2.51 | p1F | 0.0881 | 3.87 | p7V | -0.0789 | 3.48 | | | |
| m5L | 0.0573 | 3.08 | p1N | 0.0681 | 2.26 | p8I | -0.1019 | 3.87 | | | |
| m5M | 0.1460 | 4.23 | p1Q | -0.1121 | 3.1 | p8K | -0.0791 | 3.2 | | | |
| m5Q | 0.1052 | 3.43 | p1R | 0.1203 | 4.78 | p8L | -0.0503 | 2.62 | | | |
| m5T | 0.1044 | 4.05 | p1S | 0.1142 | 4.89 | p8M | -0.1475 | 3.12 | | | |
| m6D | 0.0728 | 2.93 | p1W | 0.1531 | 2.73 | p8P | 0.0770 | 2.85 | | | |
| m6E | -0.0593 | 2.49 | p2T | 0.1190 | 10.31 | p8R | -0.0660 | 2.34 | | | |

* $mi\alpha$ and $pi\alpha$ are abbreviations for $minus_{i\alpha}$ and $plus_{i\alpha}$, respectively; where $i$ is the position and $\alpha$ is an amino acid
** BH denotes Beta Hairpin

To test the selected LPM, it is applied on the chosen out-of-sample dataset. This dataset comprises of 152 proteins that are not yet curated to be N-glycosylated at certain positions. The LPM predicted that 140 of the 152 proteins have a probability less than 50% of being N-glycosylated at these positions. The distribution of these probabilities is shown in Figure 3.

**Ridge Regression Estimated Linear Probability Model Predictions of N-glycosylation in Proteins with Structural and Sequence Data**

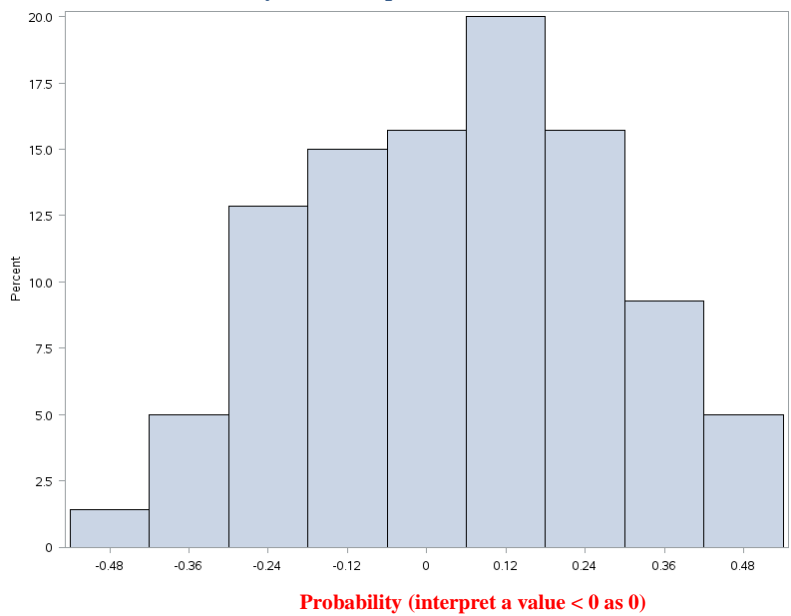

Figure 3: Distribution of N-glycosylation probabilities < 50% for not yet curated proteins

On the other hand, the LPM predicts that 12 sites of the 152 sites have probabilities greater than 50% of being N-glycosylated. The top 6 of these 12 predicted probabilities are shown in Table 4 and the resultant proteins are cross-checked against Uniprot and their protein descriptions are noted. We predict these additional sites will be N-glycosylated. Time will tell if this will turn out, experimentally, to be the case.

Table 4: Probabilities of glycosylation > 50% for not yet curated sites

| Uniprot Accession Number | Uniprot Position | LPM prediction | Protein descriptor |
|---|---|---|---|
| P01848 | 79 | 66.3% | T-cell receptor alpha chain C region |
| P33527 | 819 | 64.9% | Multidrug resistance-associated protein 1 |
| P12643 | 338 | 59.4% | Bone morphogenetic protein 2 |
| Q6P179 | 103 | 58.7% | Endoplasmic reticulum aminopeptidase 2 |
| P51124 | 46 | 56.2% | Granzyme M |
| Q15399 | 248 | 55.7% | Toll-like receptor 1 |

Because when only a single sugar is bound to an N-site it is not clear whether the protein is glycosylated, the LPM is applied to predict such proteins to see if this uncertainty in glycosylation identification is reflected in model predictions. In *glycos_public.xlsx* there are 1,339 "N-glycosylated" proteins that have only a single sugar bound to the N-site. This is an out-of-sample data subset as it has not been used to train the LPM. When the LPM is applied to this out-of-sample data subset, it is found that 716 of the 1,339 (about 53%) proteins have predicted probabilities greater than 50%. The distributions of predicted probabilities for the 716 proteins are shown in Figure 4. The distribution of predicted probabilities for the remaining 623 proteins (about 47% of the 1,339) are shown in Figure 5. The nearly 50-50 split in predicting N-glycosylation in this data subset is reflective of the uncertainty in the claim that proteins with a single sugar bound are N-glycosylated.

**Ridge Regression Estimated Linear Probability Model Predictions of N-glycosylation in Proteins with Structural and Sequence Data**

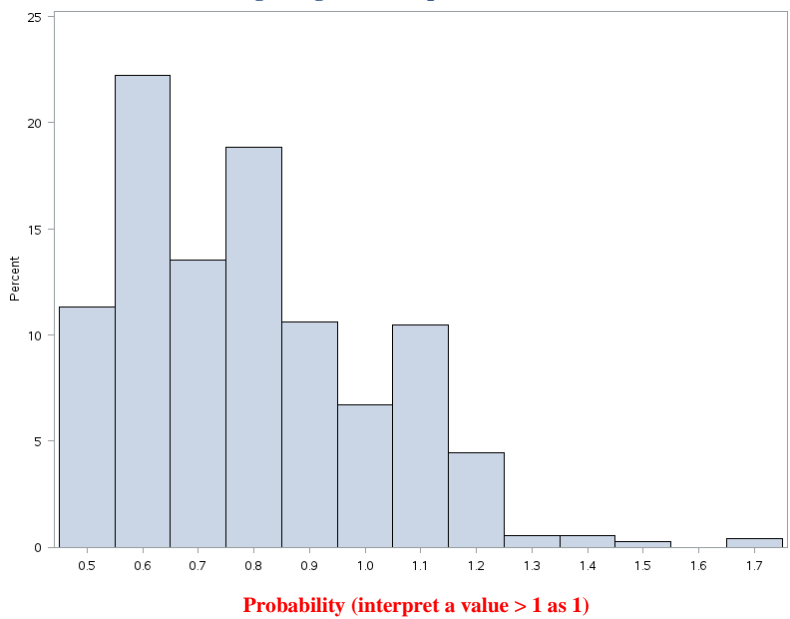

Figure 4: Distribution of N-glycosylation probabilities > 50% for single-sugar-bound proteins

**Probability (interpret a value > 1 as 1)**

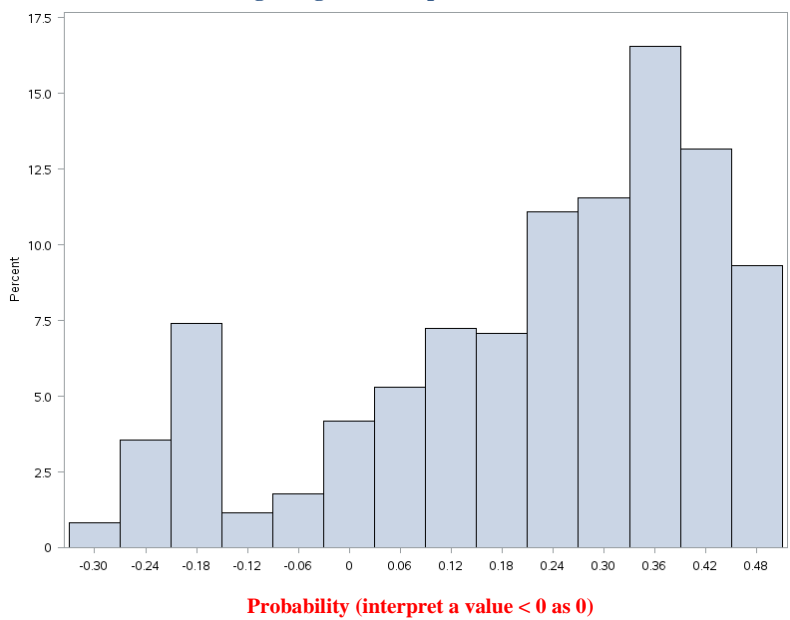

Figure 5: Distribution of N-glycosylation probabilities < 50% for single-sugar-bound proteins

**Probability (interpret a value < 0 as 0)**

Some other studies (Chuang, G.Y. et al., 2012; Hamby & Hirst, 2008) have employed "machine learning" techniques like random forests (Breiman, 2001a) to predict N-glycosylation. However, this approach was not pursued in this paper because the focus herein is on the explanatory process driving N-glycosylation. Thus, classical hypothesis testing is germane to the present work, and algorithmic modeling with its pervasive focus on prediction is not, in this paper, an end in itself[15]. In this sense, Cox's crucial comment to Breiman (Breiman, 2001b) is echoed, which is that the "starting point" in this paper is not the data, but the underlying process generating it[16]. Furthermore, echoing Cox, again, the preference in this paper is to avoid proceeding with "a directly empirical black-box approach" in favor of trying to "take account of some underlying explanatory process" that can be simply represented by the unfashionable[17] LPM.

### Predicting N-Glycosylation given a Gain-of-N-Site

The LPM is applied to a set of 10 human sequences retrieved from the curated section of the Uniprot database (www.uniprot.org) and published literature that had at least one gain of glycosylation site as a result of a single nucleotide polymorphism (Mazumder, Morampudi, Motwani, & Vasudevan, 2012). Gain of glycosylation sites have been shown to have deleterious effects leading to diseases (Vogt, G. et al., 2007). The 10 sequences are ranked by predicted probabilities of glycosylation. These sequences may yield interesting biological insights. Two of the 10 sequences are selected as case studies and presented below. While several of the proteins, among the selected sequences, are already known glycoproteins, there are a few that do not qualify to be glycoproteins with lack of a signal peptide. These may be false positives given by the model; or interesting details about them may emerge as we understand the process better. The collected data on these sequences is shown in Table 5.

**Table 5: Sequences having gained an N-site**

| Uniprot Accession No. | Uniprot Position of the N-Site | PDB ID | Protein Length | The single nucleotide polymorphism position in **bold red** results in a gain of N-site | ASA | Secondary Structure | Turn Type | Phi | Psi |
|---|---|---|---|---|---|---|---|---|---|
| P01008 | 77 | 3KCG | 464 | GSEQKIPEATNR**S**VWELSKAN | 0.44 | Loop | | -73.3 | 122.2 |
| P01009 | 341 | 4PYW | 418 | GITKVFSNGA**N**LSGVTEEAPL | 0.36 | Gamma turn | Inverse | -82.0 | 70.5 |
| P04062 | 227 | 2NT0 | 536 | PWTSPTWLKTNG**T**VNGKGSLK | 0.26 | Beta turn | I | -100 | 1.8 |
| P06280 | 408 | 3HG3 | 429 | EWTSRLRSHIN**T**TGTVLLQLE | 0.31 | Beta turn | II | -60.1 | 145.3 |
| P08236 | 379 | 3HN3 | 651 | DFNLLRWLGANA**S**RTSHYPYA | 0.04 | Loop | | -106 | 7.0 |
| P08670 | 306 | 3TRT | 466 | ADLSEAANRNND**S**LRQAKQES | 0.54 | Helix | | -64.1 | -31.4 |
| P15848 | 384 | 1FSU | 533 | FDVWKTISEG**N**PSPRIELLHN | 0.20 | Loop | | -74.9 | 160.4 |
| P34059 | 310 | 4FDI | 522 | GGSNGPFLCG**N**QTTFEGGMRE | 0.12 | Beta turn | II | -43.1 | 143.5 |
| P51688 | 86 | 4MHX | 502 | SLLTGLPQHQNG**T**YGLHQDVH | 0 | Helix | | -88.1 | -15.7 |
| Q14749 | 81 | 1R74 | 295 | DSIMLVEEGF**N**VTSVDASDKM | 0.54 | Beta strand | | -100 | 90.3 |

---

[15] As Brad Efron (Breiman, 2001b) notes: *Leo's paper is at its best when presenting the successes of algorithmic modeling, which comes across as a positive development for both statistical practice and theoretical innovation. This isn't an argument against traditional data modeling any more than splines are an argument against polynomials. The whole point of science is to open up black boxes, understand their insides, and build better boxes for the purposes of mankind.*

[16] In particular, Sir David Cox notes: *The absolutely crucial issue in serious mainstream statistics is the choice of a model that will translate key subject-matter questions into a form for analysis and interpretation. If a simple standard model is adequate to answer the subject matter question, this is fine: there are severe hidden penalties for overelaboration.*

[17] Although the LPM is unfashionable these days, its merits are independent of that perception. Here we echo Efron (Breiman, 2001b): *... New methods always look better than old ones. Neural nets are better than logistic regression, support vector machines are better than neural nets, etc. In fact it is very difficult to run an honest simulation comparison, and easy to inadvertently cheat by choosing favorable examples, or by not putting as much effort into optimizing the dull old standard as the exciting new challenger ... Complicated methods are harder to criticize than simple ones. By now it is easy to check the efficiency of a logistic regression, but it is no small matter to analyze the limitations of a support vector machine ...*

**Ridge Regression Estimated Linear Probability Model Predictions of N-glycosylation in Proteins with Structural and Sequence Data**

For the data in Table 5, the LPM predicts that the two highest probabilities of N-glycosylation are for sequences P51688 and P04062 with probabilities of about 53% and 45%, respectively. Although the predicted probability for P04062 is less than 50%, it is discussed because its predicted probability falls within one standard deviation (27%) of the mean predicted probability (71%) of N-glycosylation for the glycosylated human sequences. The predicted probabilities of N-glycosylation for the remaining 8 sequences in Table 5 are much lower than 50%.

*Case 1*: **N-sulphoglucosamine sulphohydrolase** (SGSH_HUMAN)

Loss in the activity of SGSH has been associated with lysosomal storage disease mucopolysaccharidosis IIIA, an inherited metabolic defect known as MPS III A and Sanfilippo syndrome A (Sidhu, N.S. et al., 2014; Fiorentino, F. et al., 2006). This disease is characterized by the accumulation of heparin sulfate due to inactivity in the degradation pathway. Several mutations (about 100) have been reported. One of the variations reported is the mutation of an M to T at Uniprot position 88 that the LPM has identified as a potential glycosylation site gained as a result of the mutation. A homology model, using Swiss-pdb viewer (Guex, Peitsch, & Schwede, 2009), of this variant was created using the available structure (PDB-ID: 4MHX) of SPHM as a template. This enzyme has been shown to exist as a dimer and higher order oligomers. This site is in close proximity to the dimer interface of the enzyme as shown in Figure 6, which was created using PyMOL. Based on homology modeling, it is predicted that its effect can cause a severe phenotype, disrupting the dimer interface due to steric hindrance when a sugar is bound. This indicates a major conformational change. Further studies are needed to validate these predictions.

**Figure 6: Cartoon representation of N-sulphoglucosamine sulphohydrolase\***

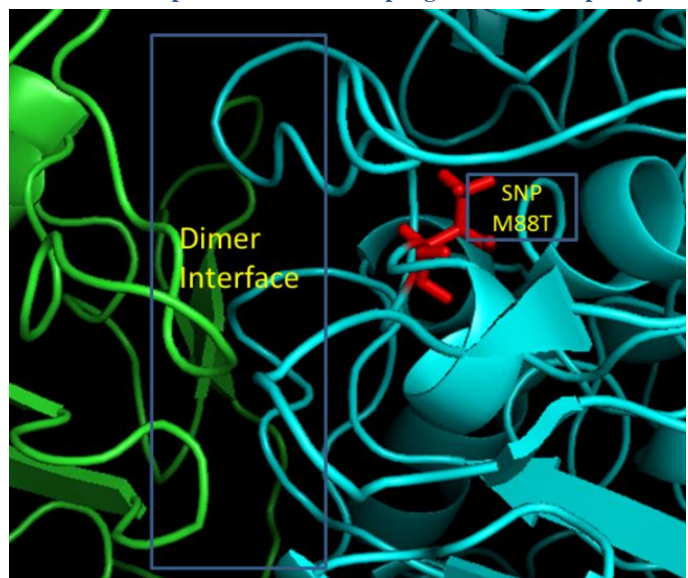

\* The N-site is shown as sticks in red

*Case 2*: Glucosylceramidase *(GLCM_HUMAN)*

Beta-glucocerebrosidase is a housekeeping enzyme that helps break down glucocerebroside into a sugar (glucose) and a simpler fat molecule (ceramide).

Glucocerebroside is a component of the membrane and gets broken down when cells die and recycled as new cells are formed. It forms an integral component of the lysosomes and any defect in this enzyme due to polymorphisms leads to Gaucher Disease, Parkinsons disease and Dementia with Lewy bodies (Neumann, J. et al., 2009). Several mutations have been identified in this gene. Here a polymorphism at position A229T leading to gain of an N-Site is discussed. This mutation has been seen in patients with Type 1 and II Gaucher disease (Koprivica, V. et al., 2000; Lieberman, R.L. et al., 2007). This site is at the tetramer-interface and likely affects the function of the enzyme (Figure 7, which was created in PyMOL). Further studies are needed to validate these predictions.

**Figure 7: Cartoon representation of Glucosylceramidase\***

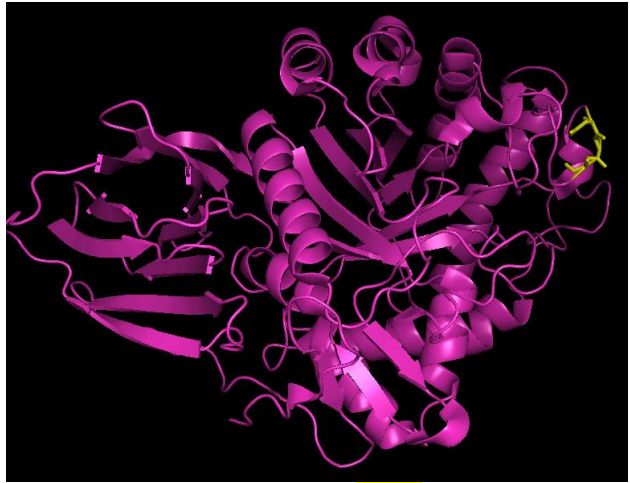

\* The N-site is shown as sticks in yellow

## Concluding Remarks

This paper has demonstrated that the ridge regression estimated LPM can be used to predict N-glycosylation. Over time, the sequences predicted, by the LPM, to be N-glycosylated can be tracked to monitor whether or not they turn out to be experimentally valid. The efficacy of the LPM in predicting N-glycosylation can be compared and contrasted with other prediction methodologies. The use of machine learning as an explanatory variable selection procedure to specify the LPM may be worth exploring.

**Ridge Regression Estimated Linear Probability Model Predictions of N-glycosylation in Proteins with Structural and Sequence Data**

# Ridge Regression Estimated Linear Probability Model Predictions of N-glycosylation in Proteins with Structural and Sequence Data

# Ridge Regression Estimated Linear Probability Model Predictions of N-glycosylation in Proteins with Structural and Sequence Data